\documentclass[11pt]{article}
\usepackage{amsmath,amssymb,epsf}

\def\ben{\begin{equation}}
\def\een{\end{equation}}

  \let\n=\nu

\let\C=\Chi

\def\nn{\nonumber} \def\bd{\begin{document}} \def\ed{\end{document}}
\def\ds{\documentstyle} \let\fr=\frac \let\bl=\bigl \let\br=\bigr
\let\Br=\Bigr \let\Bl=\Bigl
\let\bm=\bibitem
\let\na=\nabla
\let\pa=\partial \let\ov=\overline
\newcommand{\be}{\begin{equation}}
\newcommand{\ee}{\end{equation}}
\def\ba{\begin{array}}
\def\ea{\end{array}}
\def\ft#1#2{{\textstyle{\frac{\scriptstyle #1}{\scriptstyle #2}}}}
\def\fft#1#2{\frac{#1}{#2}}
\def\del{\partial}
\def\vp{\varphi}
\def\sst#1{{\scriptscriptstyle #1}}
\def\oneone{\rlap 1\mkern4mu{\rm l}}
\def\td{\tilde}
\def\wtd{\widetilde}
\def\ie{\rm i.e.\ }
\def\dalemb#1#2{{\vbox{\hrule height .#2pt
        \hbox{\vrule width.#2pt height#1pt \kern#1pt
                \vrule width.#2pt}
        \hrule height.#2pt}}}
\def\square{\mathord{\dalemb{6.8}{7}\hbox{\hskip1pt}}}
\newcommand{\ho}[1]{$\, ^{#1}$}
\newcommand{\hoch}[1]{$\, ^{#1}$}
\newcommand{\bea}{\begin{eqnarray}}
\newcommand{\eea}{\end{eqnarray}}
\newcommand{\ra}{\rightarrow}
\newcommand{\lra}{\longrightarrow}
\newcommand{\Lra}{\Leftrightarrow}
\newcommand{\ap}{\alpha^\prime}
\newcommand{\bp}{\tilde \beta^\prime}
\newcommand{\tr}{{\rm tr} }
\newcommand{\Tr}{{\rm Tr} }
\def\0{{\sst{(0)}}}
\def\1{{\sst{(1)}}}
\def\2{{\sst{(2)}}}
\def\3{{\sst{(3)}}}
\def\4{{\sst{(4)}}}
\def\5{{\sst{(5)}}}
\def\6{{\sst{(6)}}}
\def\7{{\sst{(7)}}}
\def\8{{\sst{(8)}}}
\def\n{{\sst{(n)}}}
\def\cA{{{\cal A}}}
\def\cB{{{\cal B}}}
\def\cF{{{\cal F}}}
\def\tV{\widetilde V}
\def\tW{\widetilde W}
\def\tH{\widetilde H}
\def\tE{\widetilde E}
\def\tF{\widetilde F}
\def\tA{\widetilde A}
\def\im{{{\rm i}}}
\def\tY{{{\wtd Y}}}
\def\ep{{\epsilon}}
\def\vep{{\varepsilon}}
\def\R{\rlap{\rm I}\mkern3mu{\rm R}}
\def\bD{{{\bar D}}}

\def\R{\rlap{\rm I}\mkern3mu{\rm R}}
\def\bD{{{\bar D}}}
\def\R{{{\mathbb R}}}
\def\C{{{\mathbb C}}}
\def\H{{{\mathbb H}}}
\def\CP{{{\mathbb C}{\mathbb P}}}
\def\RP{{{\mathbb R}{\mathbb P}}}
\def\Z{{{\mathbb Z}}}
\def\bA{{{\mathbb A}}}
\def\bB{{{\mathbb B}}}
\def\bC{{{\mathbb C}}}
\def\bD{{{\mathbb D}}}
\def\bE{{{\mathbb E}}}
\def\bZ{{{\mathbb Z}}}
\def\Re{{{\frak{Re}}}}
\def\Im{{{\frak{Im}}}}
\def\cosec{{\,\hbox{cosec}\,}}
\def\Gm{{\Gamma_{\!\! -}}}
\def\Gp{{\Gamma_{\!\! +}}}
\def\stan{{standard }}
\def\nonstan{{supernumerary }}

\begin{document}
\begin{flushright}
MIFP-04-16 \\
{\bf hep-th/0408143}\\
August\  2004
\end{flushright}

\begin{center}
 
{\large {\bf PP-waves in AdS Gauged Supergravities\\
and Supernumerary
Supersymmetry}}
 
\vspace{20pt}

J. Kerimo and H. L\"u\hoch{1}

\vspace{20pt}

{\it George P. \&  Cynthia W. Mitchell Institute for Fundamental
Physics,\\
Texas A\&M University, College Station, TX 77843-4242, USA}

\vspace{40pt}
 
\underline{ABSTRACT}
\end{center}

   Purely gravitational pp-waves in AdS backgrounds are described by
the generalised Kaigorodov metrics, and they generically
preserve $\ft14$ of the maximum supersymmetry allowed by the AdS spacetimes.
We obtain $\ft12$ supersymmetric purely gravitational pp-wave solutions,
in which the Kaigorodov component is set to zero.  We construct
pp-waves in AdS gauged supergravities supported by a vector field. 
We find that the solutions in $D=4$ and $D=5$ can then preserve
$\ft12$ of the supersymmetry.  Like those in ungauged supergravities,
the supernumerary supersymmetry imposes additional constraints on the
harmonic function associated with the pp-waves.  These new backgrounds
provide interesting novel features of the supersymmetry enhancement
for the dual conformal field theory in the infinite-momentum frame.

{\vfill\leftline{}\vfill \vskip 10pt \footnoterule {\footnotesize
\hoch{1} Research supported in part by DOE grant
DE-FG03-95ER40917.
}


\newpage

\section{Introduction}

     Maximally supersymmetric pp-waves of type IIB \cite{bfhp} and
M-theory \cite{kg} provide simple but non-trivial backgrounds for
studying \cite{bmn} the AdS/CFT correspondence \cite{malda,gkp,wit}
since string theory on such a background becomes massive and exact
solvable \cite{metsaev}.  These solutions can also be obtained from
the Penrose limit of AdS$\times$sphere backgrounds of the
corresponding theories, and thus are supported by form field strengths
instead of being purely gravitational.  A large class of pp-waves
supported by field strengths in M-theory and type-IIB supergravities
were studied in \cite{clp1,clp2,gh}.  These solutions in general have
16 ``standard'' Killing spinors, that is half of the maximum
supersymmetry.  For appropriate choices of field strengths and
integration constants, supernumerary Killing spinors beyond the 16
standard ones could also arise \cite{clp1,clp2,gh,bena,mich1,mich2}.  
These include all of those from the Penrose limit of AdS$\times$sphere
arising from non-dilatonic p-branes and/or intersecting p-branes, and
of AdS$\times$sphere$\times$sphere, arising from non-standard brane
intersections \cite{lpv}.

        It is natural to study the pp-waves in AdS background.  The
effect of introducing a pp-wave can be viewed as performing an
infinite boost on the boundary conformal field theory
\cite{clpboost,bcr}.  The purely gravitational pp-wave in AdS$_4$ was
constructed in the 1960s and has been known as the Kaigorodov metric
\cite{kaig}.  Higher dimensional generalisations, namely, the purely
gravitational pp-waves in higher AdS spacetimes, were obtained in
\cite{clpboost}.  These metrics are homogeneous and preserve
$\ft14$ of the supersymmetry,
consisting with the fact that in the dual conformal field theory, the
original supersymmetry as well as the superconformal symmetry, are
broken by the boost \cite{clpboost}.  The Kaigorodov metrics can
also be generalised to a class of inhomogeneous solutions, obtained
in \cite{siklos,ozsvath,gibbonsruback}.

          In this paper, we construct pp-waves in AdS gauged
supergravities that are not only purely gravitational like Kaigorodov 
metrics but are also those supported in addition by a field strength of
the theories as well.  We show by explicit construction that supernumerary
supersymmetry can arise with appropriately chosen field strength and
integration constants in $D=4$ and $D=5$.  The new solutions preserve
$\ft12$ of the supersymmetry, double the number of standard Killing
spinors associated with the general pp-wave solutions including the
Kaigorodov metric.  In fact we show that even in the case of pure
gravitational pp-waves, supernumerary supersymmetry can arise, extending
the result of \cite{bcr}, where only $\ft14$ supersymmetric
purely gravitational pp-waves were discussed.  The arising of the
supernumerary supersymmetry provides novel features of supersymmetry
enhancement for the dual conformal field theories in the
infinite-momentum frame.

        The paper is organised as follows:  In section 2,
we discuss the supersymmetry of the purely gravitational pp-waves
in the  Einstein theory with a cosmological constant in arbitrary
dimensions.  We show that $\ft12$ supersymmetric solutions can arise.
In section 3, we obtain
pp-waves in $D=4$, ${\cal N}=2$ Einstein-Maxwell gauged supergravity
supported by the Maxwell field.  We obtain explicit supernumerary
Killing spinors as well as the standard Killing spinors.  In section
4, we obtain the analogous solutions in $D=5$, ${\cal N}=2$
Einstein-Maxwell gauged supergravity.  We show that supernumerary
Killing spinors also arise in this case.  In section 5, we obtain
$U(1)$-charged pp-wave solutions to minimal AdS gauged supergravities
in $D=6$ and $D=7$.  In these two cases, the solutions have only
standard Killing spinors but no supernumerary ones.  We conclude our
paper in section 5.  In appendix A, we uplift some of our solutions to
M/string theories.  In appendix B, we present a general class of
pp-waves supported by an $n$-form field strength in a $D$-dimensional
AdS gravity theory.

\section{Purely gravitational pp-waves}

      In this section, we consider purely gravitational pp-waves in
Einstein gravity with a negative cosmological constant in arbitrary dimensions.  
The Lagrangian is given by
\be
e^{-1}{\cal L}=R+(D-1)(D-2)g^2,
\ee
where $e={\big(}\! -\mbox{det}(g_{\sst{MN}}){\big)}^{1/2}$.
The Killing spinor in this theory satisfies the equation
\be
\nabla_{\sst M}\epsilon=-\ft12g\, \Gamma_{\sst M}\epsilon\,.
\ee
We study AdS pp-waves using the metric ansatz
\be
ds_{\sst D}^2=e^{2g\rho}(-4dx^+dx^- + H(dx^+)^2 + dz_i\, dz_i) + d\rho^2,
\ee
where the function $H$ depends on $x^+$, $\rho$ and $z_i$ coordinates.
The Einstein equations of motion reduce to
\be
\square H \equiv \Big(\del_\rho^2 + g (D-1)\del_\rho +
e^{2g\rho}\sum_{i=1}^{D-3} \del^2_i\Big) H=0\,,\label{gendeom}
\ee
where the index $i$ stands here for $z_i$.
To discuss the Killing spinor equations, we make a natural choice for
the vielbein basis
\be
e^+=e^{g\rho} dx^+\,,\quad
e^-=e^{g\rho} (-2dx^- + \ft12 H\, dx^+)\,,\quad
e^i = e^{g\rho} dz^i\,,\quad e^\rho=d\rho\,,
\ee
such that we have $ds^2 = 2 e^+e^- + e^ze^z + e^\rho e^\rho$.
In this tangent basis, the spin connections are given by
\be
\omega_{-\rho} = g\, e^+\,,\quad
\omega_{+i} =\ft12 e^{-g\rho}\, \del_i H\, e^+\,,\quad
\omega_{+\rho} =g\, e^{-} + \ft12 H'\, e^+\,,\quad
\omega_{i\rho} =g\, e^i\,,\label{gendspincon}
\ee
where the prime denotes the derivative $\del_{\rho}$\,.
Note that for the metric in this basis we have $\eta_{+-}=1$ and 
$\eta_{++}=\eta_{--}=0$. In the following we use the notation 
that all derivatives are with respect to the curved metric and 
all indices on gamma matrices are vielbein indices.
The Killing spinor equations are given by
\bea
&&[\del_{+}+\ft12ge^{g\rho}\,\Gamma_+(\Gamma_\rho+1)
+\ft14ge^{g\rho} H\, \Gamma_-(\Gamma_\rho+1)
+\ft14e^{g\rho} H'\,\Gamma_{-\rho} + \ft14\sum_i^{D-3}
\del_i H\, \Gamma_{-i}]\epsilon=0\,,\nn\\
&&[\del_{-}-ge^{g\rho}\,\Gamma_-(\Gamma_\rho+1)]\epsilon=0\,,\nn\\
&&[\del_i+\ft12ge^{g\rho}\,\Gamma_i(\Gamma_\rho+1)]\epsilon=0\,, \qquad
i=1,2,\cdots, D-3,\nn\\
&&[\del_{\rho} + \ft12g\,\Gamma_\rho]\epsilon=0\,,
\eea
where we have $\Gamma_+^2 = \Gamma_-^2 =0$ and $\{\Gamma_+,
\Gamma_-\}=2$.  
Thus, we see that a generic pp-wave in a pure Einstein theory with
a cosmological constant preserves $\ft14$ of the maximally allowed
supersymmetry.  The projections are given by
\be
(\Gamma_\rho + 1)\epsilon=0=\Gamma_-\epsilon\,.
\ee

     We are interested in finding solutions that preserve more
supersymmetry.  One might expect that it would be helpful in this case
first to analyse the integrability 
conditions $[\del_{\sst M}\,,\del_{\sst N}]\epsilon=0$ among the Killing 
spinor equations. This calculation yields
\bea
0&=&[\del_+\,,\del_i]\epsilon= - \ft14{\big[}ge^{2g\rho}H'\, 
\Gamma_i + e^{g\rho}\del_i H'\,\Gamma_{\rho} 
+ \sum_j \del_j\del_i H\, \Gamma_j{\big]} \Gamma_- \epsilon\,,\nn\\
0&=&[\del_+\,,\del_{\rho}]\epsilon=-\ft14{\big[}e^{g\rho}(H'' + 2gH')
\Gamma_{\rho} + \sum_i\del_i H'\,\Gamma_i{\big]}\Gamma_-\epsilon\,.
\eea
The integrability conditions are satisfied provided that
$\Gamma_-\epsilon=0$.  This is an example where integrability
conditions are not enough for the existence of the Killing spinors.

      To see whether the metrics can admit more supersymmetry than
the $\ft14$\,, let us use the less restrictive projection condition
\be
g(\Gamma_{\rho} + 1)\epsilon={\rm i} f\,\Gamma_-\epsilon\,,
\label{adsppproj}
\ee
where $f=f(x^+,\rho,z_i)$ is to be determined.   Substituting
this projection into the Killing spinor equations, we have
\bea
&&{\Big[}\del_+ + \ft{\rm i}2e^{g\rho}f\,\Gamma_+\, \Gamma_- 
- \ft14{\Big(}e^{g\rho} H' + \sum_i
\Gamma_i \del_i H{\Big)}\Gamma_- {\Big]}\epsilon=0\,,\nn\\
&&\del_-\,\epsilon=0\,,\qquad
[\del_i + \ft{\rm i}2e^{g\rho} f\,\Gamma_i\, \Gamma_-]\,\epsilon=0\,,
\qquad
{\big[}\del_{\rho} + \ft{\rm i}2 f\, \Gamma_- - \ft12g{\big]}
\epsilon=0\,.\label{gspinor}
\eea
The integrability conditions $[\del_{\sst M}\,,
\del_{\sst N}]\epsilon=0$ among these equations are
\bea
0&=&[\del_i\,,\del_j]\epsilon=-\ft{\rm i}2e^{g\rho}(\Gamma_j\,\del_i f 
- \Gamma_i\,\del_j f)\Gamma_-\epsilon\,,\nn\\
0&=&[\del_i\,,\del_{\rho}]\epsilon=\ft{\rm i}2[(e^{g\rho} f)'\,\Gamma_i
- \del_i f]\Gamma_-\epsilon\,,\nn\\
0&=&[\del_+\,,\del_i]\epsilon=
-\ft12{\Big[}{\rm i}e^{g\rho}(\Gamma_i\, \del_+ f
- \Gamma_+\, \del_i f) + e^{2g\rho} f^2\,\Gamma_i
+ \ft12e^{g\rho}\del_i H'\nn\\
&&\qquad + \ft12\sum_{j=1}^{D-3}\Gamma_j\del_j\del_i H {\Big]}
\Gamma_-\epsilon\,,\nn\\
0&=&[\del_+\,,\del_\rho]\epsilon=-\ft12{\Big[}{\rm i}\del_+ f 
+ e^{g\rho} f^2 - {\rm i}(e^{g\rho} f)'\,\Gamma_+
+ \ft12 \sum_i\Gamma_i\del_i H'\nn\\
&&\qquad + \ft12e^{g\rho}(H''+gH'){\Big]}\Gamma_-\epsilon\,.
\eea
From these integrability conditions we see that if we insist on more
supersymmetry than the usual $\ft14$ we must set
\be
\del_i f = 0 = \del_i H' \qquad
\hbox{and}\qquad
\del_i\del_j H = 0\,, \qquad i\neq j\,.\label{hcon1}
\ee
We then have
\bea
(e^{g\rho}f)'&=&0\,,\label{eqf1}\\
{\rm i}\del_+ f + e^{g\rho}f^2
+ \ft12 e^{-g\rho}\del_i\del_i H& = &0\,,\qquad i=1,2,\cdots,D-3\,,
\label{eqf2}\\
{\rm i}\del_+ f + e^{g\rho}f^2 +\ft12e^{g\rho}
(H''+gH')&=&0\,.\label{eqf3}
\eea
The conditions in (\ref{hcon1}), together with (\ref{gendeom}), 
imply that $H$ is given by
\bea
H=\ft12 \sum_{i=1}^{D-3} c_i z_i^2 + \fft{e^{-2g\rho}}{2g^2(D-3)}
\sum_{i=1}^{D-3} c_i + b\, e^{-(D-1)g\rho}\,,\label{nocharge}
\eea
where $c_i$ and $b$ are functions depending on $x^+$ only.
Equation (\ref{eqf2}) implies that all $c_i$'s are equal, and hence
we let $c_i=c(x^+)$.   From 
eqs.(\ref{eqf1}) and (\ref{eqf3}) it follows that we must set $b=0$. 
It is straightforward to solve for $f$, given by
\be
f=e^{-g\rho} U(x^+)\,,
\ee
where $U$ satisfies the following first-order non-linear equation
\be
\im \fft{dU}{dx^+} + U^2 + \ft12 c=0\,.\label{Sd4eq}
\ee
Making use of eq.(\ref{Sd4eq}) together with the solutions for $f$ and $H$ 
we can now solve the Killing spinor equations given in (\ref{gspinor}). 
The Killing spinor solution is
\bea
\epsilon &=& e^{\ft12g\rho} \Big(1 -\ft{\im}2 U
\sum_{i=1}^{D-3} z_i\, \Gamma_{i}\, \Gamma_-\Big)
\Big(1 + \ft{\rm i}{2} g^{-1} f\, \Gamma_-\Big)\nn\\
&&\quad \times {\Big[}1-\ft12{\Big(}1-e^{-\im\int U dx^+}{\Big)}
\Gamma_+\, \Gamma_-{\Big]}\epsilon_0\,,\label{adsppks}
\eea
where $\epsilon_0$ is a constant spinor satisfying 
$(\Gamma_{\rho}+1)\epsilon_0 = 0$.
Thus, the metric preserves $\ft12$ of the supersymmetry.  It is important
that the final result of our Killing spinors (\ref{adsppks}) satisfy
the projection condition (\ref{adsppproj}), which can be easily verified
to be true.

    Note that the special case of $c=0$, $b\ne 0$ is the
Kaigorodov metric.  The above analysis implies that it preserves
$\ft14$ of the supersymmetry.  In order to have $\ft12$ BPS solutions,
it is necessary to set the Kaigorodov component to zero.

       Note that in general $c$ is any function depending on $x^+$.
The simplest case is that $c$ is a constant.   The $x^+$ dependence
of $c$ has no effect on the existence of the Killing spinors, but only
modifies the explicit Killing spinor solutions.

\section{PP-waves in $D=4$ gauged supergravity}

\subsection{The solution}

In this section we continue our investigations of supernumerary 
supersymmetry by including a U(1) charge.
We start with gauged ${\cal N}=2$ Einstein-Maxwell AdS
supergravity, whose Lagrangian for the bosonic sector is given by
\be
e^{-1}{\cal L}_4 = R - \ft14 F_\2^2 + 6 g^2,
\ee
where $F_\2=dA_\1$.
The supersymmetry transformation rule for the complex gravitino
$\Psi_{\sst M}=\Psi_{\sst M}^1+{\rm i}\Psi_{\sst M}^2$ is
\cite{freed,frad}
\be
\delta \Psi_{\sst M} = \Bigl[\nabla_{\sst M} -
\ft{\rm i}{2}g A_{\sst M} +
\ft{\rm i}{8} F_{\sst{AB}}\, \Gamma^{\sst{AB}}{}\,\Gamma_{\sst M}
+ \ft12 g\, \Gamma_{\sst M}\Bigr]\epsilon\,.
\ee

    We consider the following pp-wave ansatz
\bea
ds^2 &=& e^{2g\rho} (-4dx^+\, dx^- + H (dx^+)^2 + dz^2) +
d\rho^2,\nn\\
A_\1 &=& g^{-1}S(1-e^{-g\rho})\, dx^+,
\label{ansatz4}
\eea
where $H = H(x^+,\rho,z)$ and $S$ is here a function of $x^+$.
The equations of motion imply that $H$ satisfies
\be
\square H \equiv H''+ 3g\, H'+e^{-2g\rho}\del_z^2 H = -S^2e^{-4g\rho}\,.
\label{ads4ppwave}
\ee
The solution can be expressed as
\be
H=S^2 \Bigl(\ft12c\, z^2 + g^{-2}\, (\ft12 c\,
e^{-2g \rho} - \ft14 e^{-4g\rho} + b\, e^{-3g\rho})\Bigr) +
 H_0\,,\label{d4gensol}
\ee
where $b$ and $c$ are functions of $x^+$ and $H_0$ satisfies
$\Box H_0=0$.  (Note that the terms associated with $b$ and $c$
actually belong to $H_0$\,.  We extract them since they are necessary for
the solution to reduce under $g\rightarrow 0$ to the pp-wave that is
the Penrose limit of AdS$_2\times S^2$ of the corresponding ungauged
theory.)

         If we turn off the field strength by setting $S=0$, and let
$H_0$ depend only on $\rho$, namely $H_0=c_0 + b\, e^{-3g\rho}$, then
we recover the Kaigorodov metric.

\subsection{Standard supersymmetry}

      Here we investigate the supersymmetry of the
``charged'' pp-wave we derived.  The Killing spinor equations in this
background are given by
\bea
&&[\del_++\ft12ge^{g\rho}\,\Gamma_+(\Gamma_\rho+1)
+\ft14ge^{g\rho}H\, \Gamma_-(\Gamma_\rho+1)
+\ft14\Gamma_{-z}\del_z H + \ft14e^{g\rho}H'\,\Gamma_{-\rho}\nn\\
&&\qquad +\ft{\rm i}2S(e^{-g\rho}-1)
+\ft{\rm i}4e^{-g\rho}S\,\Gamma_\rho\,\Gamma_-\, \Gamma_+]
\epsilon=0\,,\nn\\
&&[\del_--ge^{g\rho}\,\Gamma_-(\Gamma_\rho+1)]\epsilon=0\,,\nn\\
&&[\del_z+\ft12ge^{g\rho}\,\Gamma_z(\Gamma_\rho+1)
+ \ft{\rm i}4e^{-g\rho}S\,\Gamma_{z\rho}\,\Gamma_-]\epsilon=0\,,\nn\\
&&[\del_{\rho}-\ft{\rm i}4e^{-2g\rho}S\,\Gamma_-
+ \ft12g\,\Gamma_\rho]\epsilon=0\,.\label{d4kseq}
\eea
Imposing the following projections
\be
(\Gamma_\rho + 1)\epsilon=0\,,\qquad \Gamma_-\epsilon=0\,,\label{genpro}
\ee
the Killing spinor equations become
\be
[\del_+ - \ft{\rm i}2S]\epsilon=0\,,
\qquad \del_-\,\epsilon=0\,,
\qquad \del_z\,\epsilon=0\,, \qquad
[\del_{\rho}-\ft12 g]\epsilon=0\,.
\ee
Thus the Killing spinor is given by
\be
\epsilon = e^{\ft12 g\rho + \ft{\im}2\!\int\! S\, dx^+}\epsilon_0\,,
\label{spinorstand4}
\ee
where $\epsilon_0$ is a constant spinor satisfying
$(\Gamma_\rho + 1)\epsilon_0=0$ and $\Gamma_-\epsilon_0=0$.  The solution
therefore preserves $\ft14$ of the supersymmetry.  We follow the
literature \cite{clp1,clp2} and call these spinors the standard
Killing spinors, since there is no further requirement on the function
$H$ for the existence of the $\epsilon$, as long as $H$ satisfies the
equation of motion (\ref{ads4ppwave}).

\subsection{Supernumerary supersymmetry}

       When the integration constants of $H$ satisfy further
conditions, there can arise additional Killing spinors, which are
called supernumerary Killing spinors in \cite{clp1,clp2}.  In order to
obtain these Killing spinors, we consider the integrability conditions
$[\del_{\sst M}\,, \del_{\sst N}]\epsilon=0$.  We find that
\bea
0&=&[\del_z\,,\,\,\del_{\rho}]\epsilon =
\ft{\rm i}4ge^{-g\rho}S\,\Gamma_z\,\Gamma_-(\Gamma_\rho+1)
\epsilon\,,\nn\\
0&=&[\del_+\,,\del_-]\epsilon =-\ft{\rm i}2gS\,
\Gamma_-(\Gamma_\rho+1)\epsilon\,,\nn\\
0&=&[\del_+\,,\del_z]\epsilon =
\ft{\rm i}4gS(3-2\Gamma_+\,\Gamma_-)\Gamma_z
(\Gamma_\rho+1)\epsilon - \ft{\rm i}4e^{-g\rho}\del_+ S\,
\Gamma_{z\rho}\,\Gamma_- \epsilon\nn\\
&& - \ft14e^{g\rho}\del_z H'\,\Gamma_{\rho}\, \Gamma_-\epsilon
- \ft14{\big[}ge^{2g\rho}H' + \del_z^2 H 
+ \ft12e^{-2g\rho}S^2{\big]}\Gamma_z\,\Gamma_-\epsilon\,,\nn\\
0&=&[\del_+\,,\del_{\rho}]\epsilon =-\ft{\rm i}4
ge^{-g\rho}S(3-\Gamma_+\,\Gamma_-)(\Gamma_\rho+1)\epsilon
+ \ft{\rm i}4e^{-2g\rho}\del_+ S\,\Gamma_-\epsilon\nn\\
&&\qquad - \ft14\del_z H'\, \Gamma_z\, \Gamma_-\epsilon
+\ft14e^{g\rho}{\big[}gH'+e^{-2g\rho}\del_z^2 H
+ \ft12e^{-4g\rho}S^2{\big]}\Gamma_\rho\,\Gamma_-\epsilon\,.
\eea
To arrive at the last integrability condition we made use of equation
(\ref{ads4ppwave}) for $H$.  It is clear that the
integrability conditions are satisfied with the projections given in
(\ref{genpro}).  However, we now show that it is possible to relax
these projections.  We find that the integrability 
conditions can also be satisfied, with the following less
restrictive projection
\be
g(\Gamma_\rho + 1)\epsilon={\rm i}f\,\Gamma_-\epsilon\,,
\label{epsproj4}
\ee
where $f=f(x^+,\rho,z)$\,. This gives the projected Killing spinor equations
\bea
&&[\del_+ - \ft{\rm i}2S-\ft12g^{-1}e^{-g\rho} f S\,\Gamma_-
+ \ft{\rm i}2(e^{g\rho} f + \ft12e^{-g\rho}S)\Gamma_+\, \Gamma_-\nn\\
&&\qquad - \ft14(e^{g\rho}H' 
+ \Gamma_z\del_z H)\Gamma_-]\epsilon=0\,, \qquad
\del_-\, \epsilon=0\,,\nn\\
&&[\del_z + \ft{\rm i}2(e^{g\rho} f
+ \ft12e^{-g\rho}S)\Gamma_z\, \Gamma_-]\epsilon=0\,,\nn\\
&&[\del_{\rho}+\ft{\rm i}2(f - \ft12e^{-2g\rho}S)\Gamma_-
- \ft12g]\epsilon=0\,. \label{pspinors4}
\eea
The integrability conditions among these equations are
\bea
0&=&[\del_z\,,\del_{\rho}]\epsilon=-\ft{\rm i}2{\big[}\Gamma_z\del_z f
- (e^{g\rho} f)' + \ft12g e^{-g\rho} S{\big]}\Gamma_z\,
\Gamma_-\epsilon\,,\nn\\
0&=&[\del_+\,,\del_z]\epsilon=-\ft12{\big[}{\rm i}
(e^{g\rho}\del_+ f + \ft12e^{-g\rho}\del_+ S)\Gamma_z
-({\rm i}e^{g\rho}\,\Gamma_+ - g^{-1}e^{-g\rho}S)\del_z f\nn\\
&&\quad +(e^{g\rho} f + \ft12e^{-g\rho}S)^2\,\Gamma_z
+\ft12(e^{g\rho}\del_z H' + \Gamma_z\del_z^2 H){\big]}
\Gamma_-\epsilon\,,\nn\\
0&=&[\del_+\,,\del_{\rho}]\epsilon=-\ft12{\big[}{\rm i}(\del_+ f
- \ft12e^{-2g\rho}\del_+ S)+g^{-1}S(e^{-g\rho} f)'\nn\\
&&\quad -{\rm i}{\big(} (e^{g\rho} f)'
- \ft12g e^{-g\rho}S{\big)}\Gamma_+
+ \ft12 e^{g\rho}(H''+gH') + \ft12 \Gamma_z\del_z H'\nn\\
&&\quad + e^{g\rho}(f^2 - \ft14e^{-4g\rho}S^2){\big]}
\Gamma_-\epsilon\,.
\eea
It is clear from these expressions that if we want more supersymmetry
than $\ft14$ we need again to impose $\del_z f = 0 = \del_z H'$.
The vanishing of the integrability conditions in this case then yields 
the equations
\bea
&&(e^{g\rho} f)'-\ft12g e^{-g\rho}S=0\,,\nn\\
&&{\rm i}(\del_+ f + \ft12e^{-2g\rho}\del_+ S)
+ \ft12e^{-g\rho}\del_z^2 H + e^{g\rho}(f
+ \ft12e^{-2g\rho}S)^2=0\,,\\
&&{\rm i}(\del_+ f - \ft12 e^{-2g\rho}\del_+ S)
+ g^{-1}S(e^{-g\rho}f)' +\ft12 e^{g\rho}(H''+gH') + e^{g\rho} f^2
- \ft14 e^{-3g\rho}S^2=0\,.\nn\label{eqc4}
\eea
From the first of eqs.(\ref{eqc4}) we obtain
\be
f=-\ft12e^{-2g\rho}S + e^{-g\rho}U\,,\label{f4}
\ee
where $U=U(x^+)$ is in general a complex function.
Note that $S$ is a real function. Using the solution for $f$
and the equation for $H$ the remaining two equations in (\ref{eqc4}) gives
\bea
&&{\rm i}\fft{dU}{dx^+} + U^2 + \ft12\del_z^2H = 0\,,\nn\\
&&{\rm i}\fft{dS}{dx^+} - e^{-g\rho}S^2 + 3S\, U
+ ge^{3g\rho}H' + e^{g\rho}\del_z^2 H = 0\,.\label{system4}
\eea
Since the functions $S$ and $U$ depend only on $x^+$ we need to check
that the $\rho$ dependence in the equation for $S$ drops out
before we can proceed. For this we need to make use of
the solution for $H$, which is given by (\ref{d4gensol}).  Setting
$H_0=0$, and substituting $H$ into eqs.(\ref{system4}) we have\footnote{
It is straightforward to verify that in general supernumeary supersymmetry
requires that $H_0$ be given by (\ref{nocharge}), which is not the
most general solution for $\Box H_0=0$\,.}
\be
{\rm i}\fft{dS}{dx^+} - 3S(b\, S - U) = 0\,, \qquad
{\rm i}\fft{dU}{dx^+} + U^2 + \ft12c\,S^2 = 0\,. \label{system4f}
\ee
In order to solve these equations we rewrite $U$ into an real and imaginary
part $U=u+{\rm i}v$. Eqs.(\ref{system4f}) then yield the following
set of equations:
\bea
&&\fft{dS}{dx^+}+3v\,S=0\,, \qquad \fft{du}{dx^+}+2u\,v=0\,,\qquad
S(u-b\,S)=0\,,\nn\\
&&\fft{dv}{dx^+}+v^2-u^2-\ft12c\, S^2=0\,.\label{eqSi}
\eea
We have four equations for the five functions $S,u,v,b$ and $c$\,, 
and so one function will be left arbitrary.
We present the solution to eqs.(\ref{eqSi}) in terms of 
the function $b$.  The solution is given by
\bea
S&=&\fft{k}{b^3}\,, \qquad u=\fft{k}{b^2}\,,\qquad 
v=b^{-1}\fft{db}{dx^+}\,,\nn\\
c&=&\fft{2b^5}{k^2}{\Big[} \fft{d^2b}{dx^{+2}}
- \fft{k^2}{b^3}{\Big]}\,,\label{d4xxxx1}
\eea
where $k$ is an arbitrary constant and we have taken $S \neq 0$.
(The case with $S=0$ was considered in section 2.) Note
that the original generic $\ft14$ supersymmetric solution depending on
the three functions $b$, $c$ and $S$ now only have one independent
function in order for the solution to have the
enhanced $\ft12$ supersymmetry.

     We  next turn to presenting the explicit Killing spinors.
The Killing spinor equations are
\bea
&&[\del_+ - \ft{\rm i}2S - \ft12g^{-1}e^{-g\rho}f S\,\Gamma_-
+ \ft{\rm i}2 U\,\Gamma_+\, \Gamma_-
- \ft14(e^{g\rho}H' + c\,z\,S^2\, \Gamma_z)\Gamma_-]\epsilon=0\,,\nn\\
&&\del_-\,\epsilon=0\,, \qquad 
[\del_z + \ft{\rm i}2 U\, \Gamma_z\, \Gamma_-]\epsilon=0\,, \qquad
[\del_{\rho} - \ft{\rm i}2g^{-1}f'\, \Gamma_- - \ft12g]\epsilon=0\,,
\label{pspinors4f}
\eea
where $f$ is given by (\ref{f4}).
The third equation of the above implies $\epsilon=(1 - \ft{\im}{2}
z\, U\, \Gamma_z\, \Gamma_-)\times\,\chi(\rho,x^+)$.
Substituting this into the fourth equation yields the solution 
$\chi=e^{\ft12 g\rho}(1 + \ft{\im}{2} g^{-1}f\, \Gamma_-)\eta(x^+)$\,.  
The equation for $\eta$ can be obtained from the first equation of 
(\ref{pspinors4f}) after making use of eqs.(\ref{system4f}).
We have 
\be
\fft{d\eta}{dx^+}-\ft{\rm i}2[S - U\,\Gamma_+\, \Gamma_-]\eta = 0\,.
\ee
Note that it requires conspiracy for the $z$ and $\rho$ dependent
terms to drop out.  Finally, we arrive at the Killing spinor, given by
\bea
\epsilon&=&e^{\ft12g\rho +\ft{\rm i}2\! \int\! S\, dx^+}
(1-\ft{\rm i}2 z\, U\,\Gamma_z\, \Gamma_-)
(1+\ft{\rm i}2g^{-1}f\,\Gamma_-)\nn\\
&&\quad \times{\Big[}1-\ft12(1-e^{-{\rm i} \int U dx^+})
\Gamma_+\, \Gamma_-{\Big]}\epsilon_0\,,\label{spinorg4}
\eea
where $\epsilon_0$ is a constant spinor, satisfying the projection
\be
(\Gamma_\rho + 1) \epsilon_0=0\,.
\ee

    There are two special cases that are worth considering.  The first case
is that $b$ is set to a constant, implying that $v=0$.  It follows then that
the functions $S$ and $u$ are constants as well, and $c=-2b^2$. 
Assuming $S=\mu$ the Killing spinor in this case is given by
\bea
\epsilon&=&e^{\ft12g\rho +\ft{\rm i}2 \mu x^+}
(1-\ft{\rm i}2\mu b\, z\, \Gamma_z\, \Gamma_-)
(1+\ft{\rm i}2g^{-1}f\,\Gamma_-)\nn\\
&&\quad \times{\Big[}1-\ft12(1-e^{-{\rm i}\, \mu b x^+})
\Gamma_+\, \Gamma_-{\Big]}\epsilon_0\,,\label{spinor4}
\eea
where $\epsilon_0$ is a constant spinor, satisfying the projection
$(\Gamma_\rho + 1) \epsilon_0=0$.
Thus after imposing the condition $c=-2b^2$, the solution has
$\ft12$ of the supersymmetry instead of the $\ft14$ for a generic
pp-wave solution.  The standard Killing spinors are those with an
additional projection $\Gamma_-\epsilon_0=0$, in which case,
$\epsilon$ of (\ref{spinor4}) becomes that in (\ref{spinorstand4}).
The supernumerary Killing spinors are the remaining half with
$\Gamma_-\epsilon_0 \ne 0$.

     The function $H$, for the pp-wave with supernumerary
supersymmetry, is given by
\bea
H &=& -\mu^2\,b^2z^2 - g^{-2}f^2 =  -\mu^2{\Big(}b^2 z^2
+ g^{-2}(b^2\,e^{-2g\rho}+ \ft14e^{-4g\rho}
- b\,e^{-3g\rho}){\Big)}\,,\nn\\
f &=& -\ft12\mu(e^{-2g\rho}-2b\, e^{-g\rho})\,.
\eea
If we set $b=\ft12$\,, we have $H=-\mu^2\, [\ft14 z^2 + g^{-2}
\sinh^2(\ft12 g\rho)\, e^{-3g\rho}]$.  We can then take the
$g\rightarrow 0$ limit and obtain a pp-wave in ungauged
$D=4$, ${\cal N}=2$ Einstein Maxwell supergravity.  The solution
is given by
\bea
ds^2&=&-4dx^+\, dx^- -\ft14 \mu^2 (z^2 + \rho^2)\, (dx^+)^2 +
dz^2 + d\rho^2,\\
F_\2&=& -\mu\, dx^+\wedge d\rho\,.
\eea
This is precisely the pp-wave arising from the Penrose limit of
AdS$_2\times S^2$, which is known to have supernumerary
supersymmetries \cite{clp1,clp2}.
                                                                                
        Note that in the ansatz (\ref{ansatz4}), we could instead have
used $A_\1 = \mu z\, dx^+$.  The metric in this case is identical to
that with $A_\1$ given in (\ref{ansatz4}).  However, we verified that
the solution would be non-supersymmetric, because of the explicit
$A_\1$ dependence in the supersymmetry transformation rule.
                                                                                
      Charged pp-waves with $c=0$ were also obtained in \cite{cai},
by performing an infinite boost of the AdS charged black holes.  It
can be deduced from the above analysis that the solution with $c=0$
has only the standard supersymmetry.  We can also obtain pure
gravitational $\ft12$ supersymmetric pp-waves by setting $b=\td b/\mu$
and then sending $\mu\rightarrow 0$.
                                                                                
       In \cite{klemm1} a general class of pp-waves that preserve
$\ft14$ of the supersymmetry were given.  PP-waves with $\ft12$ of the
supersymmetry were also obtained in \cite{klemm2}, where the Killing
spinors were given in component language, whilst ours are presented in
an elegant form, in terms of constant spinors satisfying a single
gamma matrix projection.

       The second special case corresponds to the absence of 
the Kaigorodov component $b$ which can be achieved by taking 
a degenerate limit of (\ref{d4xxxx1}).  
It is worth examing on its own.  In this case we have the coupled system
\be
\fft{dS}{dx^+} + 3v\,S = 0\,, \qquad
\fft{dv}{dx^+} + v^2 - \ft12c\,S^2 = 0\,.\label{eqSv4}
\ee
This implies a relation between the metric functions $c$ and $S$, given by
\be
c=-\ft23 S^{-3}\fft{d^2S}{{dx^+}^2} 
+ \ft89 S^{-4}{\Big(}\fft{dS}{dx^+}{\Big)}^2.
\ee
Making use of these equations together with the solutions for $H$ and $f$ the
Killing spinor equations (\ref{pspinors4f}) yield the solution
\bea
\epsilon&=&e^{\ft12g\rho}\,e^{\ft{\rm i}2 \! \int \! S\, dx^+}
(1+\ft12z\,v\,\Gamma_z\, \Gamma_-)(1+\ft{\rm i}2g^{-1}f\,\Gamma_-)\nn\\
&&\quad \times {\Big[}1-\ft12(1-e^{\int \! v\, dx^+})
\Gamma_+\, \Gamma_-{\Big]}\epsilon_0\,,
\eea
where $\epsilon_0$ is a constant spinor satisfying
$(\Gamma_{\rho}+1)\epsilon_0=0$. For the functions $H$ and $f$
we have
\bea
H&=&\ft12S^2{\Big[}c\, z^2 + \ft12g^{-2}e^{-2g\rho}
(2c - e^{-2g\rho}){\Big]}\,,\nn\\
f&=&-\ft12e^{-2g\rho}S + {\rm i}e^{-g\rho}v\,.
\eea
We can consider a special case of eqs.(\ref{eqSv4}) by setting 
$c \equiv$\;constant and $v=\td k S$ where $\td k$ is a (real) constant.
In this case the equations fixes $\td k$ to $\td k^2=-\ft14c$
with $c < 0$.  The equation for $S$ is
\be
\fft{dS}{dx^+}+\td k S^2=0\,,
\ee
with the solution given by $S(x^+)=1/(1+\td k\,x^+)$\,.

\section{PP-waves in $D=5$ gauged supergravity}

\subsection{The solution}

         For simplicity, we consider simple gauged supergravity in
$D=5$.  The Lagrangian for the bosonic sector is given by \cite{mg}
\be
e^{-1}{\cal L}_5=R-\ft14F_\2^2+
\ft1{12\sqrt3}\epsilon^{\sst{MNPQR}}F_{\sst{MN}}
F_{\sst{PQ}}A_{\sst R}+12g^2.
\ee
Analogous to the $D=4$ discussion, we use the following pp-wave ansatz
\bea
ds^2 &=& e^{2g\rho} (-4dx^+\, dx^- + H (dx^+)^2 + dz_1^2 + dz_2^2) +
d\rho^2,\nn\\
A_\1 &=& \ft12 g^{-1}S(1-e^{-2g\rho})\, dx^+,
\label{ansatz5}
\eea
where $S=S(x^+)$.
The supergravity equations of motion then reduce to the following
\be
\square H\equiv H''+4gH'+ e^{-2g\rho}\,
\sum_{i=1}^2\del_i^2H =-
e^{-6g\rho}S^2.\label{d5heom}
\ee
The solution is given by
\be
H=S^2{\big[}\ft12(c_1 z_1^2+ c_2 z_2^2)+g^{-2}(\ft14(c_1 + c_2)
\,e^{-2g\rho} -\ft1{12}e^{-6g\rho}+b\,e^{-4g\rho}){\big]} + H_0\,,
\label{d5gensol}
\ee
where $c_i$ and $b$ are functions of $x^+$ and $\Box H_0=0$.  
The generalised Kaigorodov-type metric is obtained
by setting $S=0$ and $H_0=c_0 + b\, e^{-4g\rho}$ with $c_0$ and $b$ 
now being constants.

\subsection{Supersymmetry}

The supersymmetry transformation on the gravitino is given by
\be
\delta\Psi_{\sst M}=[\nabla_{\sst M}-\ft{3\,\rm i}{2\sqrt3}\,gA_{\sst M}
-\ft{\rm i}{16\sqrt3}F_{\sst {AB}}\,(\Gamma_{\sst M}\, \Gamma^{\sst{AB}}
-3\,\Gamma^{\sst{AB}}\,\Gamma_{\sst M})+
\ft12g\,\Gamma_{\sst M}]\epsilon\,,
\ee
where $\epsilon$ is a complex symplectic spinor.  For our pp-wave
background, the Killing spinor equations are given by
\bea
&&[\del_++\ft12ge^{g\rho}\,\Gamma_+(\Gamma_\rho+1)
+\ft14ge^{g\rho}H\,\Gamma_-(\Gamma_\rho+1)
+\ft14e^{g\rho}H'\,\Gamma_{-\rho}\nn\\
&&\qquad + \ft14\sum_{i=1}^2\Gamma_{-i}\del_i H
+\ft{3\,\rm i}{4\sqrt3}S(e^{-2g\rho}-1)\nn\\
&&\qquad +\ft{\rm i}{8\sqrt3}e^{-2g\rho}S\,\Gamma_\rho
(\Gamma_+\,\Gamma_- + 3\Gamma_-\,\Gamma_+)]\epsilon=0\,,\nn\\
&&[\del_- - ge^{g\rho}\,\Gamma_-(\Gamma_\rho+1)]\epsilon=0\,,\nn\\
&&[\del_i+\ft12ge^{g\rho}\,\Gamma_i(\Gamma_\rho+1)
+\ft{\rm i}{4\sqrt3} e^{-2g\rho}S\,\Gamma_{i\rho}\,\Gamma_-]
\epsilon=0\,, \qquad i=1,2,\nn\\
&&[\del_{\rho}-\ft{\rm i}{2\sqrt3} e^{-3g\rho}S\,
\Gamma_- + \ft12g\,\Gamma_\rho]\epsilon=0\,.
\label{d5originalks}
\eea
As in the case of $D=4$, the standard Killing spinors, which exist for
all $H$ satisfying (\ref{d5heom}), arise with the following projections
$(\Gamma_\rho + 1)\epsilon=0$ and $\Gamma_-\epsilon=0$.  The Killing
spinor equations become
\be
[\del_+ - \im\ft{\sqrt3}4 S]\epsilon=0\,,
\qquad \del_-\,\epsilon=0\,,
\qquad \del_i\,\epsilon=0\,, \qquad
[\del_{\rho}-\ft12 g]\epsilon=0\,.
\ee
Thus, the generic pp-waves we considered preserve $\ft14$ of the
standard supersymmetry.  In \cite{gg}, a general class of null
solutions with $\ft14$ of the supersymmetry were obtained, however,
the issue of supernumerary supersymmetry was not addressed.  We
demonstrate below that, as in the case of $D=4$, supernumerary
Killing spinors can also arise.

        To obtain the supernumerary Killing spinor and the
corresponding conditions on $H$, we impose the following projection
on the spinors
\be
g(\Gamma_{\rho} + 1)\epsilon = {\rm i} f\, \Gamma_-\epsilon\,.
\ee
The Killing spinor equations become
\bea
&&{\big[}\del_+ - \ft{3\rm i}{4\sqrt3}S + \ft{\rm i}2 (e^{g\rho} f
+\ft1{2\sqrt3}e^{-2g\rho} S)\Gamma_+\, \Gamma_-
-\ft14\sum_i\Gamma_i\, \Gamma_-\del_i H\nn\\
&&\qquad \quad -\ft14(e^{g\rho}H'+\sqrt3g^{-1}
e^{-2g\rho} f\, S)\Gamma_-{\big]}\epsilon=0\,, \qquad
\del_-\, \epsilon=0\,,\nn\\
&&[\del_i + \ft{\rm i}2(e^{g\rho} f
+ \ft1{2\sqrt3}e^{-2g\rho}S)\Gamma_i\, \Gamma_-]\epsilon=0\,,\nn\\
&&[\del_{\rho}+\ft{\rm i}2(f - \ft1{\sqrt3}e^{-3g\rho}S)\Gamma_-
- \ft12g]\epsilon=0\,.
\eea
The integrability conditions among these equations are
\bea
0&=&[\del_i\,,\del_{\rho}]\epsilon=-\ft{\rm i}2{\big[}\del_i f
- (e^{g\rho} f)'\,\Gamma_i + \ft1{\sqrt3}g e^{-2g\rho}
S\,\Gamma_i{\big]}\Gamma_-\epsilon\,,\nn\\
0&=&[\del_+\,,\del_i]\epsilon=-\ft12{\big[}
{\rm i}(e^{g\rho}\del_+ f + \ft1{2\sqrt3}e^{-2g\rho}
\del_+ S)\Gamma_i - ({\rm i}e^{g\rho}\,\Gamma_+
- \ft3{2\sqrt3}g^{-1}e^{-2g\rho}S)\del_i f\nn\\
&&\quad +\ft12e^{g\rho}\del_i H'
+ \ft12\sum_j\Gamma_j\del_j\del_i H
+(e^{g\rho} f + \ft1{2\sqrt3}e^{-2g\rho}S)^2\, \Gamma_i
{\big]}\Gamma_-\epsilon\,,\nn\\
0&=&[\del_+\,,\del_{\rho}]\epsilon=-\ft12{\big[}{\rm i}(\del_+ f
- \ft1{\sqrt3}e^{-3g\rho}\del_+ S)
+ \ft3{2\sqrt3}g^{-1}S(e^{-2g\rho} f)'
+ \ft12\sum_i\Gamma_i\del_i H'\nn\\
&&\quad -{\rm i}{\big(} (e^{g\rho} f)'
- \ft1{\sqrt3}g e^{-2g\rho}S{\big)}\Gamma_+
+ \ft12e^{g\rho}(H''+gH')\nn\\
&&\quad + (f -\ft1{\sqrt3}e^{-3g\rho}S)(e^{g\rho}f
+ \ft1{2\sqrt3}e^{-2g\rho}S){\big]}\Gamma_-\epsilon\,.
\eea
To have more supersymmetry than the $\ft14$ we need to set
\be
\del_i f = 0 = \del_i H' \quad \mbox{and} \quad
\del_j\del_i H = 0\,, \quad i \neq j\,.
\ee
The integrability conditions then imply
\bea
&&f=-\ft1{2\sqrt3}e^{-3g\rho}S + e^{-g\rho}U\,,\nn\\
&&{\rm i}\fft{dU}{dx^+} + U^2 + \ft12\del_i^2H=0\,,\qquad
i=1,2,\nn\\
&&{\rm i}{\Big(}\fft{dS}{dx^+}-\ft2{\sqrt3}e^{2g\rho}\fft{dU}
{dx^+}{\Big)} - g^{-1}e^{3g\rho}S(e^{-2g\rho}f)'
- \ft1{\sqrt3}e^{4g\rho}(H''+gH')\nn\\
&&\qquad \qquad \qquad \qquad \quad -\ft2{\sqrt3}e^{3g\rho}U
(f-\ft1{\sqrt3}e^{-3g\rho}S)=0\,,
\eea
where $U=U(x^+)$.  Substituting in the solution for $H$, given by
(\ref{d5gensol}), we find that it is necessary to have that $c_1=c_2\equiv c$,
and that $H_0$ is given by (\ref{nocharge}).  For simplicity, we set $H_0=0$
here since the $H_0$ represents the pure gravitational component,
which was discussed in section 2. The equations for $S$ and $U$ are
then given by
\be
{\rm i}\fft{dS}{dx^+}-4S(\sqrt3\,b\,S - U)=0\,, \qquad
{\rm i}\fft{dU}{dx^+} + U^2 + \ft12c\,S^2=0\,.
\ee
Substituting $U = u + {\rm i}v$ into the above
yields the equations
\bea
&&\fft{dS}{dx^+}+4v\,S=0\,, \qquad \fft{du}{dx^+}+2u\,v=0\,, \qquad
S(u-\sqrt3\,b\,S)=0\,,\nn\\
&&\fft{dv}{dx^+} + v^2 - u^2 - \ft12c\, S^2 = 0\,.\label{system5}
\eea
The solution to these equations is
\bea
S&=&\fft{k}{b^2}\,, \qquad u=\fft{\sqrt3\, k}{b}\,, \qquad
v=\fft1{2b}\fft{db}{dx^+}\,,\nn\\
c&=&\fft{b^3}{k^2}{\Big[}\fft{d^2b}{dx^{+2}}
- \fft1{2b}{\Big(}\fft{db}{dx^+}{\Big)}^{\! 2}\, {\Big]} - 6\, b^2,
\eea
where $k$ is an arbitrary constant and we have taken $S \neq 0$.
Note that as in the case of $D=4$,
the original generic $\ft14$-supersymmetric metric depending on the
four functions $b$, $c_1$, $c_2$ and $S$ now only have one independent
function in order for the solution to have the
enhanced $\ft12$ supersymmetry.

    The Killing spinor is calculated from the equations
\bea
&&[\del_+ - \ft{3\, \im}{4\sqrt3}S 
- \ft3{4\sqrt3}g^{-1}e^{-2g\rho}f S\,\Gamma_-
+ \ft{\im}2 U\,\Gamma_+\, \Gamma_-\nn\\
&&\qquad - \ft14{\big(}e^{g\rho}H' 
+ c\, S^2 (z_1\, \Gamma_1 + z_2\, \Gamma_2){\big)}
\Gamma_-]\epsilon=0\,,\nn\\
&&\del_-\,\epsilon=0\,, \qquad 
[\del_i + \ft{\rm i}2 U\, \Gamma_i\, \Gamma_-]\epsilon=0\,, \qquad
[\del_{\rho} - \ft{\im}2g^{-1}f'\, \Gamma_- - \ft12g]\epsilon=0\,.
\label{pspinors5f}
\eea
The solution is
\bea
\epsilon&=&e^{\ft12g\rho +\im\ft{\sqrt3}4 \! \int\! S\, dx^+}
\big(1-\ft{\rm i}2 U\,(z_1\, \Gamma_1 + z_2\, \Gamma_2)\Gamma_-\big)
(1+\ft{\im}2g^{-1}f\,\Gamma_-)\nn\\
&&\quad \times{\Big[}1-\ft12(1-e^{-\im \int U dx^+})
\Gamma_+\, \Gamma_-{\Big]}\epsilon_0\,,\label{spinorg5}
\eea
where $\epsilon_0$ is a constant spinor satisfying 
$(\Gamma_{\rho}+1)\epsilon_0=0$.  
As in $D=4$ we consider two special cases.  The first corresponds
to $v=0$\,, which implies that $b$, $c$ and $S$ are all constants,
with $c=-6b^2$.  Letting $S=\mu$ the Killing spinor in this case is 
given by
\bea
\epsilon &=& e^{\ft12g\rho + \im \ft{\sqrt3}{4}\mu x^+}
\Big(1 - \im \ft{\sqrt3}2\,\mu b(z_1\, \Gamma_1 + z_2\, \Gamma_2)\,
\Gamma_-\Big)(1 + \ft{\im}{2}g^{-1}f\, \Gamma_-)\nn\\
&&\qquad \times \Bigl[1 - \ft12(1-e^{-\im\sqrt3\,\mu b x^+})
\Gamma_+\, \Gamma_-\Bigr]\epsilon_0\,,
\eea
where $\epsilon_0$ is a constant spinor satisfying $(\Gamma_\rho +1)
\epsilon_0=0$. Thus the solution preserves
half of the supersymmetry.
Among all the Killing spinors, the standard ones are
those with $\Gamma_-\epsilon_0=0$, whilst the remaining half with
$\Gamma_-\epsilon_0 \ne 0$ are the supernumerary ones. 
         The function $H$ for the pp-waves with supernumerary
supersymmetry is given by
\bea
H &=& -3\mu^2 b^2(z_1^2+z_2^2)-g^{-2}f^2\nn\\
&=&-\mu^2[3b^2(z_1^2+z_2^2)+g^{-2}(3b^2\,e^{-2g\rho}
+\ft1{12}e^{-6g\rho}-b\,e^{-4g\rho})]\,,\nn\\
f &=& -\ft1{2\sqrt3}\mu(e^{-3g\rho}-6b\, e^{-g\rho})\,.
\eea
If we further let $b=\ft16$\,, we have $H=-\ft1{12}\mu^2 (z_1^2 + z_2^2
+ 4 g^{-2} \sinh^2(g\rho)\, e^{-4g\rho})$.  This enables us to take
the limit $g\rightarrow 0$, giving rise to a pp-wave in the
corresponding ungauged $D=5$ supergravity, given by
\bea
ds^2 &=& -4 dx^+dx^- - \ft1{12}\mu^2\,(z_1^2 + z_2^2 + 4\rho^2)\,
(dx^+)^2 +dz_1^2 + dz_2^2 + d\rho^2,\nn\\
F_\2 &=& -\mu dx^+\wedge d\rho\,.
\eea
This pp-wave can also arise from the Penrose limit of AdS$_3\times
S^2$ or AdS$_2\times S^3$, which have supernumerary supersymmetries.

   The second case is that of $b=0$, and hence eqs.(\ref{system5}) reduce to
\be
\fft{dS}{dx^+} + 4v\,S = 0\,,\qquad
\fft{dv}{dx^+} + v^2 - \ft12c\, S^2 = 0\,.\label{eqSv5}
\ee
The Killing spinor is then given by
\bea
\epsilon&=&e^{\ft12g\rho}\,e^{{\rm i}\ft{\sqrt3}4\!
\int\! S\,dx^+} {\big(}1+\ft12v(z_1\, \Gamma_1 
+ z_2\, \Gamma_2)\Gamma_-{\big)}
(1+\ft{\rm i}2g^{-1}f\,\Gamma_-)\nn\\
&&\qquad \times {\Big[}1-\ft12(1-e^{\int\! v\, dx^+})
\Gamma_+\, \Gamma_-{\Big]}\epsilon_0\,,\label{spinor5}
\eea
where $\epsilon_0$ is a constant spinor satisfying
$(\Gamma_{\rho}+1)\epsilon_0=0$ and
\bea
H&=&\ft12S^2{\Big[}c\, (z_1^2+z_2^2) + 2g^{-2}e^{-2g\rho}
(c - \ft1{12}e^{-4g\rho}){\Big]}\,,\nn\\
f&=&-\ft1{2\sqrt3}e^{-3g\rho}S + {\rm i} e^{-g\rho}v\,.
\eea
If we specialise to $v=\td kS$ and $c=-6\td k^2$
where $\td k$ is a constant, the system (\ref{eqSv5}) simplifies to
\be
\fft{dS}{dx^+} +4\td k S^2 = 0\,.
\ee

\section{PP-waves in $D=6$ and $D=7$}

\subsection{$D=6$}

      Our next example is in the Romans six-dimensional gauged ${\cal
N}=(1,1)$ supergravity \cite{rom1}. The bosonic field content
comprises the metric, a dilaton $\phi$, a $2$-form potential, a $U(1)$
potential and the gauge potentials $A^i_\1$ of $SU(2)$ Yang-Mills. The
Lagrangian describing the bosonic sector is \cite{clp5}
\bea
{\cal L}&=&R\ast\oneone - \ft12 {*d\phi} \wedge d\phi + (2g_1^2X^2
+\ft83g_1g_2X^{-2} - \ft29 g_2^2X^{-6})\ast\!\oneone\nn\\
&-&\ft12X^4\ast\! F_\3 \wedge F_\3 -
\ft12X^{-2}{\Big(}\!\!\ast\! G_\2 \wedge G_\2
+ \ast F^a_\2 \wedge F^a_\2{\Big)}\\
&-&A_\2 \wedge (\ft12dB_\1 \wedge dB_\1 + \ft13g_2A_\2 \wedge dB_\1
+ \ft2{27}g_2^2A_\2 \wedge A_\2 + \ft12F^a_\2 \wedge F^a_\2)\,,\nn
\eea
where $X\equiv e^{-\ft1{2\sqrt2}\phi}$\,, $F_\3=dA_\2$\,, $G_\2=dB_\1
+ \ft23g_2A_\2$\,, $F^a_\2=dA^a_\1 + \ft12g_1\epsilon_{abc}\,A^b_\1
\wedge A^c_\1$.  The fermions of this theory comprise
symplectic-Majorana gravitinos $\Psi_{\sst Mi}$ and dilatinos
$\lambda_i$ where $i=1,2$ is an $SP(1)$ index.  The supersymmetry
transformations are given by \cite{radion}
\bea
\delta\Psi_{\sst M i}&=&[D_{\sst M} - \ft1{48}X^2F_{\sst{ABC}}\,
\Gamma^{\sst{ABC}}\,\Gamma_{\sst M} \Gamma^7 - \ft1{4\sqrt2}
(g_1X + \ft13g_2X^{-3})\,\Gamma_{\sst M}]\epsilon_i\nn\\
&&\qquad +\ft1{16\sqrt2}(\Gamma_{\sst M}\,
\Gamma^{\sst{AB}}-2\Gamma^{\sst{AB}}\,\Gamma_{\sst M})
X^{-1}(G_{\sst{AB}}\,\delta_i{}^j - {\rm i}\,\Gamma^7
F_{\sst{AB}\,i}{}^j)\Gamma^7\epsilon_j\,,\nn\\
\delta\lambda_i&=&[-\ft1{2\sqrt2}\Gamma^{\sst M}\del_{\sst M}\phi
+ \ft1{24}X^2F_{\sst{ABC}}\,\Gamma^{\sst{ABC}}\,\Gamma^7
+ \ft1{2\sqrt2}(g_1X - g_2X^{-3})]\epsilon_i\nn\\
&&\qquad +\ft1{8\sqrt2}X^{-1}(G_{\sst{AB}}\,\delta_i{}^j
- {\rm i}\,\Gamma^7 F_{\sst{AB}\,i}{}^j)\,
\Gamma^{\sst{AB}}\,\Gamma^7 \epsilon_j\,.
\eea
The gauge covariant derivative is defined as 
$D_{\sst M}\epsilon_i=\nabla_{\sst M}\epsilon_i 
+ \ft{\rm i}2g_1A_{\sst M\,i}{}^j\epsilon_j$ 
where $A_{\sst M\,i}{}^j\equiv A_{\sst M}^a(-\sigma^a)_i{}^j$ 
with the field strength given by
$F_{\sst{MN}i}{}^j=\del_{\sst M}A_{{\sst N}i}{}^j
+ \ft{\rm i}2g_1A_{{\sst M}i}{}^k A_{{\sst N}k}{}^j
- (\sst M \leftrightarrow \sst N)$ 
and $\sigma^a$ are the usual Pauli matrices.

    In this paper, we consider pp-wave solutions supported by only one
field strength.  Owing to the Chern-Simons modifications to various
field strengths, we find that this can only be done with a $U(1)$
vector field coming from the $SU(2)$ Yang-Mills.  Thus we consistently
set all the remaining form fields to zero, and also without loss
of generality (while insisting on AdS background)
take $g_1=g_2=-3g/\sqrt2$.  This leads to the pp-wave ansatz
\bea
ds^2 &=& e^{2g\rho} (-4dx^+\, dx^- + H (dx^+)^2 + dz_1^2 + dz_2^2 +
dz_3^2) + d\rho^2,\nn\\
A_\1 &=& \ft13 g^{-1}S(1-e^{-3g\rho})\, dx^+,
\label{ansatz6}
\eea
where $S=S(x^+)$. The equations of motion reduce to
\be
\square H\equiv H'' + 5 g H' + e^{-2g \rho} \sum_{i=1}^3
\del_i^2 H=-e^{-8g\rho}S^2,
\ee
and the solution for $H$ is given by
\be
H=S^2{\Big[}\ft12\sum_{i=1}^3c_i z_i + g^{-2}{\big(}\ft16 (c_1+c_2+c_3)
e^{-2g\rho} - \ft1{24}e^{-8g\rho}+b\, e^{-5g\rho}{\big)}{\Big]} + H_0\,,
\label{h6}
\ee
where $\Box H_0=0$. The $b$ and $c_i$ are functions of $x^+$.

        We now investigate the supersymmetry of the pp-waves.  This is
more conveniently done if we rewrite the symplectic-Majorana spinors
using a Dirac notation. (See \cite{ker} for details.)  The Killing
spinor equations from the gravitino transformation rule are given by
\bea
&&{\big[}\del_+ - \ft{\rm i}{2\sqrt2}S + \ft{\rm i}2(e^{g\rho}f 
+ \ft1{4\sqrt2}e^{-3g\rho}S)\Gamma_+\, \Gamma_-
- \ft14\sum_i\Gamma_i\del_i H\, \Gamma_-\nn\\
&&\qquad -\ft1{2\sqrt2}(g^{-1}e^{-3g\rho}f S 
+ \ft1{\sqrt2}e^{g\rho}H')\Gamma_-{\big]}\epsilon=0\,, \qquad
\del_-\, \epsilon=0\,,\nn\\
&&[\del_i + \ft{\rm i}2(e^{g\rho}f + \ft1{4\sqrt2}e^{-3g\rho}S)
\Gamma_i\, \Gamma_-]\epsilon=0\,, \qquad i=1,2,3,\nn\\
&&[\del_{\rho} + \ft{\rm i}2(f - \ft3{4\sqrt2}e^{-4g\rho}S)\Gamma_-
-\ft12g]\epsilon=0\,,
\eea
where we have made use of the projection condition 
$g(\Gamma_{\rho}+1)\epsilon={\rm i}f\,\Gamma_-\epsilon$ and where
$f=f(x^+,\rho,z_i)$. The integrability conditions 
$[\del_{\sst M}\,,\del_{\sst N}]\epsilon=0$ among these projected 
Killing spinor equations are
\bea
0&=&[\del_i\,,\del_{\rho}]\epsilon =-\ft{\rm i}2{\big[}\del_i f 
- (e^{g\rho}f + \ft1{4\sqrt2}e^{-3g\rho}S)'\,\Gamma_i{\big]}
\Gamma_-\epsilon\,,\nn\\
0&=&[\del_+\,,\del_i]\epsilon =-\ft12{\big[} {\rm i}(e^{g\rho}\del_+ f
+ \ft1{4\sqrt2}e^{-3g\rho}\del_+ S)\Gamma_i
+ \ft12\sum_j\Gamma_j\del_j\del_i H\nn\\
&&\quad + \ft12e^{g\rho}\del_i H'
+ (e^{g\rho}f + \ft1{4\sqrt2}e^{-3g\rho}S)^2\,\Gamma_i
- ({\rm i} e^{g\rho}\,\Gamma_+ - \ft1{\sqrt2}g^{-1}
e^{-3g\rho}S)\del_i f {\big]}\Gamma_-\epsilon\,,\nn\\
0&=&[\del_+\,,\del_{\rho}]\epsilon =-\ft12{\big[}
{\rm i}(\del_+ f - \ft3{4\sqrt2}e^{-4g\rho}\del_+ S)
+ \ft12\sum_i\Gamma_i\del_i H'\nn\\
&&\quad -{\rm i}(e^{g\rho}f + \ft1{4\sqrt2}e^{-3g\rho}S)'\,\Gamma_+
+\ft1{\sqrt2}g^{-1}S(e^{-3g\rho}f)'\nn\\
&&\quad +(f-\ft3{4\sqrt2}e^{-4g\rho}S)
(e^{g\rho}f+\ft1{4\sqrt2}e^{-3g\rho}S)
+ \ft12e^{g\rho}(H''+gH'){\big]}\Gamma_-\epsilon\,.
\eea
As before it is required that we set
\be
\del_i f = 0 = \del_i H'\, \qquad \mbox{and} \qquad
\del_j\del_i H = 0\,, \qquad i \neq j\,,
\ee
and $c_i=c$\,. The integrability conditions yield after using
the solution for $H$ the following results
\bea
&&f = -\ft1{4\sqrt2}e^{-4g\rho}S + e^{-g\rho}\,U\,,\nn\\
&&{\rm i}\fft{dU}{dx^+} + U^2 + \ft12c\,S^2 = 0\,,\nn\\
&&{\rm i}\fft{dS}{dx^+} + \ft1{12\sqrt2}e^{-3g\rho}S
{\big[} S(7-240b\, e^{3g\rho}) + 60\sqrt2\, e^{3g\rho}\,U{\big]} = 0\,.
\eea
In the case of $S=0$, corresponding to purely gravitational waves,
discussed in section 2, the last equation is trivially satisfied.
When $S\ne0$, due to the $\rho$ dependence, we conclude that
no supersymmetry enhancement can occur here.
This is expected, since in ungauged $D=6$, ${\cal N}=(1,1)$ supergravity, 
the pp-waves supported by a 2-form field strength also have no supernumerary 
supersymmetry.  The solution does have standard supersymmetry though.  
The Killing spinor is given by
\be
\epsilon = e^{\ft12 g\rho + \ft{\im}{2\sqrt2} \int\! S\, dx^+} 
\epsilon_0\,,
\ee
where $(\Gamma_\rho +1) \epsilon_0=0=\Gamma_-\epsilon_0$.  It is easy
to verify that the Killing spinor equations associated with both the
gravitino and dilatino transformation rules are satisfied.  Thus the
solution preserves $\ft14$ of the supersymmetry.

\subsection{$D=7$}

The Lagrangian for the bosonic sector of half-maximum supergravity in
seven dimensions \cite{town} can be written as follows \cite{lp}
\bea
{\cal L}&=&R\ast\!\oneone - \ft12 {*d\phi} \wedge d\phi -
\ft12X^4\ast\! F_\4
\wedge F_\4 - \ft12X^{-2}\ast\! F_\2^a \wedge F_\2^a\nn\\
&& +\ft12F_\2^a\wedge F_\2^a\wedge A_\3 
- \ft1{2\sqrt2}g_2F_\4\wedge A_\3\nn\\
&&+ (2g_1^2 X^2 + 2g_1g_2 X^{-3} - \ft14 g_2^2 X^{-8})\ast\!\oneone\,,
\eea
where $X=e^{-\ft1{\sqrt{10}}\phi}$, $F_\4=dA_\3$ and
$F_\2^a=dA_\1^a +\ft12g_1\epsilon_{abc}\,A_\1^b \wedge
A_\1^c$.  
In addition there is a "self-duality" condition that must be 
imposed, given by
\be
X^4\ast\! F_\4=-\ft1{\sqrt2}g_2A_\3+\ft12\omega_\3\,,
\ee
where $\omega_\3$ is defined as $\omega_\3=A_\1^a \wedge
F_\2^a - \ft16g_1\,\epsilon_{abc}\,A_\1^a\wedge A_\1^b\wedge
A_\1^c$\,.  This theory has a pair of symplectic-Majorana
gravitinos $\psi_{{\sst M}i}$ and a pair of dilatinos $\lambda_i$,
where $i=1,2$ is an $SP(1)$ index.  The fermionic supersymmetry
transformations are given by \cite{radion}
\bea
\delta\psi_{{\sst M}i}&=&\nabla_{\sst M}\epsilon_i
+ \ft{\rm i}2g_1A_{{\sst M}i}{}^j\epsilon_j 
+ \ft1{960}X^2F_{\sst{ABCD}}(\Gamma_{\sst M}\, 
\Gamma^{\sst{ABCD}}+5\Gamma^{\sst{ABCD}}\, 
\Gamma_{\sst M})\epsilon_i\nn\\
&&\qquad -\ft{\rm i}{40\sqrt2}X^{-1}(3\Gamma_{\sst M}\, \Gamma^{\sst{AB}}
-5\Gamma^{\sst{AB}}\, \Gamma_{\sst M})F_{\sst{AB}i}{}^j\epsilon_j
-\ft1{5\sqrt2}(g_1X+\ft14g_2X^{-4})\Gamma_{\sst M}\epsilon_i\,,\nn\\
\delta\lambda_i&=&[-\ft1{2\sqrt2}\Gamma^{\sst M}\del_{\sst M}\phi
+ \ft1{48\sqrt5}X^2F_{\sst{ABCD}}\, \Gamma^{\sst{ABCD}}]\epsilon_i
- \ft{\rm i}{4\sqrt{10}}X^{-1}F_{\sst{AB}i}{}^j\, \Gamma^{\sst{AB}}
\epsilon_j\nn\\
&&\quad +\ft1{\sqrt{10}}(g_1X-g_2X^{-4})\epsilon_i\,,
\eea
where $A_{{\sst M}i}{}^j \equiv A_{\sst M}^a(-\sigma^a)_i{}^j$.  Owing
to the odd-dimensional self-duality condition for the $A_\3$, our
standard ansatz for the pp-wave metric does not work for $A_\3$.  We
thus consider the pp-wave supported only by the $U(1)$ subsector of
the $SU(2)$ Yang-Mills.  The pp-wave solution is given by
\bea
ds^2 &=& e^{2g\rho} (-4dx^+\, dx^- + H (dx^+)^2 + dz_1^2 + dz_2^2 +
dz_3^2 + dz_4^2) + d\rho^2,\nn\\
A_\1 &=& \ft14 g^{-1}S (1-e^{-4g\rho})\, dx^+,
\label{ansatz7}
\eea
where $S=S(x^+)$ and $H$ satisfies
\be
\square H\equiv H'' + 6 g H' + e^{-2g\rho}\sum_{i=1}^4 
\del_i^2 H=-e^{-10g\rho}S^2.
\ee
Here we have set $g_1=g_2=-2\sqrt2\, g$.  The function $H$ can be
solved, given by
\be
H=S^2{\Big[}\ft12\sum_{i=1}^4c_i z_i^2 + g^{-2}{\big(}\ft18
\sum_{i=1}^4 c_i\, e^{-2g\rho}- \ft1{40}e^{-10g\rho} +
b\,e^{-6g\rho}{\big)}{\Big]} + H_0\,,\label{h7}
\ee
with $\Box H_0=0$ and $b$ and $c_i$ are functions of $x^+$. 

          The projected Killing spinor equations from the gravitino transformation 
rule are given by
\bea
&&{\big[}\del_+ - \ft{\rm i}{2\sqrt2}S + \ft{\rm i}2(e^{g\rho}f 
+ \ft1{5\sqrt2}e^{-4g\rho}S)\Gamma_+\, \Gamma_-
- \ft14\sum_i\Gamma_i\del_i H\, \Gamma_-\nn\\
&&\qquad -\ft1{2\sqrt2}(g^{-1}e^{-4g\rho}f S 
+ \ft1{\sqrt2}e^{g\rho}H')\Gamma_-{\big]}\epsilon=0\,,\qquad
\del_-\, \epsilon=0\,,\nn\\
&&[\del_i + \ft{\rm i}2(e^{g\rho}f + \ft1{5\sqrt2}e^{-4g\rho}S)
\Gamma_i\, \Gamma_-]\epsilon=0\,, \qquad i=1,2,3,4,\nn\\
&&[\del_{\rho} + \ft{\rm i}2(f - \ft4{5\sqrt2}e^{-5g\rho}S)\Gamma_-
-\ft12g]\epsilon=0\,.
\eea
The integrability conditions
\bea
0&=&[\del_i\,,\del_{\rho}]\epsilon =-\ft{\rm i}2{\big[}\del_i f 
- (e^{g\rho}f + \ft1{5\sqrt2}e^{-4g\rho}S)'\,\Gamma_i{\big]}
\Gamma_-\epsilon\,,\nn\\
0&=&[\del_+\,,\del_i]\epsilon =-\ft12{\big[} {\rm i}(e^{g\rho}\del_+ f
+ \ft1{5\sqrt2}e^{-4g\rho}\del_+ S)\Gamma_i
+ \ft12\sum_j\Gamma_j\del_j\del_i H\nn\\
&&\quad + \ft12e^{g\rho}\del_i H'
+ (e^{g\rho}f + \ft1{5\sqrt2}e^{-4g\rho}S)^2\,\Gamma_i
- ({\rm i} e^{g\rho}\,\Gamma_+ - \ft1{\sqrt2}g^{-1}
e^{-4g\rho}S)\del_i f {\big]}\Gamma_-\epsilon\,,\nn\\
0&=&[\del_+\,,\del_{\rho}]\epsilon =-\ft12{\big[}
{\rm i}(\del_+ f - \ft4{5\sqrt2}e^{-5g\rho}\del_+ S)
+ \ft12\sum_i\Gamma_i\del_i H'\nn\\
&&\quad -{\rm i}(e^{g\rho}f + \ft1{5\sqrt2}e^{-4g\rho}S)'\,\Gamma_+
+\ft1{\sqrt2}g^{-1}S(e^{-4g\rho}f)'\nn\\
&&\quad +(f-\ft4{5\sqrt2}e^{-5g\rho}S)
(e^{g\rho}f+\ft1{5\sqrt2}e^{-4g\rho}S)
+ \ft12e^{g\rho}(H''+gH'){\big]}\Gamma_-\epsilon\,,
\eea
imply that there is no supernumerary Killing spinors in this case.
This should be expected since in $D=7$, even in ungauged
supergravities, there is no pp-wave supported by a 2-form field
strength that has supernumerary supersymmetry.  The solution does have
$\ft14$ of standard supersymmetry, with the Killing spinor given by
\be
\epsilon=e^{\ft12 g\rho + \ft{\im}{2\sqrt2} \int\! S\, dx^+}\epsilon_0\,,
\ee
where $(\Gamma_\rho +1)\epsilon_0=0=\Gamma_-\epsilon_0$\,.

\section{Conclusions}

        In this paper, we have constructed $U(1)$-charged pp-wave
solutions in AdS gauged supergravities in $4\le D\le 7$ dimensions.
Generically these solutions preserve $\ft14$ of the supersymmetry.  In
$D=4$ and $D=5$, with an appropriate choice for the integration
constants, we have shown that supernumerary supersymmetry can arise so
that the solutions instead preserve $\ft12$ of the supersymmetry.
These solutions can take a limit to become the pp-waves that are the
Penrose limits of AdS$\times$sphere of the corresponding ungauged
supergravities. In $D=6$ and $D=7$, we find that there can be no
supernumerary supersymmetry for the $U(1)$-charged pp-waves.
We also considered a general class of purely gravitational pp-waves
in Einstein gravity with a cosmological constant in diverse
dimensions.  We showed that supernumerary supersymmetry could arise
and obtained explicitly the $\ft12$-BPS gravitational pp-waves.

     The introduction of a pp-wave in the AdS background can be viewed
as performing an infinite boost in the strong coupled dual conformal
field theory with a finite momentum density.  The non-vanishing
momentum breaks the original supersymmetry and superconformal
symmetry, and hence the supersymmetry is now $\ft14$ of the unboosted
theory.  We have shown in the supergravity side that the supersymmetry
can be doubled when the pp-wave is $U(1)$ charged, corresponding to an
$R$ charge in the dual field theory.  This indicates a novel
supersymmetry enhancement associated with the $R$ charges in the dual
three- and four-dimensional field theories.  It is of interest to
discover such a phenomenon in the dual quantum field theory in the
infinite-momentum frame.

\section*{Acknowledgments}

      We are grateful to Mirjam Cveti\v c and Chris Pope for useful
discussions.

\vskip0.5truein
\centerline{\bf \Large APPENDICES}

\vskip20pt

\appendix

\section{Uplifting to M/String theory}

   In this appendix we uplift the supersymmetric solutions supported by 
the U(1) charge to ten and eleven dimensions. In the case of the
four- and five-dimensional solutions we uplift those with $S$ being a constant.
The four- and seven-dimensional solutions are embedded in M-theory and 
the solutions in $D=5$ and $D=6$ are uplifted to type-IIB supergravity and 
to Romans massive theory, respectively.

\subsection{$D=4$ oxidised to $D=11$}

   The embedding formulae to eleven dimensions were obtained in
\cite{clp3} or we can also use the ansatz in \cite{clp4} after
truncating to our case. We obtain
\bea
d\hat s_{11}&=&e^{2g\rho}[-4dx^+dx^- - \mu^2(\ft14z^2
+ g^{-2}\sinh^2(\ft12g\rho)e^{-3g\rho})(dx^+)^2+dz^2]+d\rho^2\nn\\
&&\qquad +4 g^{-2}d\xi + g^{-2}[c^2\,(\sigma_1^2+\sigma_2^2+h_3^2)
+ s^2\,(\tilde\sigma_1^2+\tilde\sigma_2^2+{\tilde h}_3^2)]\,,\nn\\
\hat F_\4&=&-6ge^{3g\rho}dx^+\wedge dx^-\wedge d\rho\wedge dz
- \mu g^{-2}[s\,c\,d\xi\wedge \sigma_3
+ \ft12c^2\sigma_1\wedge\sigma_2\nn\\
&&\qquad - s\,c\,d\xi\wedge\tilde \sigma_3 + \ft12 s^2\,\tilde\sigma_1
\wedge\tilde\sigma_2]\wedge dx^+\wedge dz\,,
\eea
where $\sigma_i$ are the three left-invariant $1$-forms on $S^3$ 
satisfying $d\sigma_i=-\ft12\epsilon_{ijk}\sigma_j\wedge\sigma_k$\,.
They are given by $\sigma_1+{\rm i}\sigma_2=e^{-{\rm i}\psi}
(d\theta+{\rm i}\sin\theta\, d\varphi)$ and $\sigma_3=d\psi
+ \cos\theta\,d\varphi$ in terms of the Euler angles.
The $\tilde\sigma_i$ are left-invariant $1$-forms on a second $S^3$.
We have also defined
\bea
&&c\equiv\cos\xi\,, \qquad s\equiv\sin\xi\,,\nn\\
&&h_3\equiv\sigma_3-\ft12\mu(1-e^{-g\rho})dx^+, \qquad 
\tilde h_3\equiv\tilde\sigma_3-\ft12\mu(1-e^{-g\rho})dx^+,\nn\\
&&\epsilon_\3=\sigma_1 \wedge \sigma_2 \wedge h_3\,, \qquad
\tilde\epsilon_\3=\tilde\sigma_1 \wedge \tilde\sigma_2 
\wedge \tilde h_3\,.
\eea
In this pp-wave, the internal $S^7$ is twisted but not flattened.
Analogous solution but with untwisted round $S^7$ can be found in
\cite{kumar}.

\subsection{$D=5$ oxidised to type IIB}

    Using the uplifting formulae to type $IIB$ in \cite{clp3,tuan} we
obtain for the metric
\bea
&&d\hat s_{10}^2=e^{2g\rho} [-4dx^+dx^- 
- \ft1{12}\mu^2(z_1^2+z_2^2 + 4g^{-2}\sinh^2(g\rho)\,
e^{-4g\rho})(dx^+)^2\nn\\
&&\qquad + dz_1^2+dz_2^2]+ d\rho^2 + g^{-2}\sum_{i=1}^3{\big[}d\mu_i^2
+ \mu_i^2(d\phi_i + \ft1{2\sqrt3}\mu(1-e^{-2g\rho})dx^+)^2\,{\big]}\,,
\eea
and for the $5$-form field strength $F_{(5)}=G_{(5)}+\ast G_{(5)}$,
\bea
G_\5&=&-8ge^{4g\rho}dx^+\wedge dx^-\wedge d\rho\wedge d^2z
-\ft1{2\sqrt3}\mu g^{-2}\sum_{i=1}^3d(\mu_i^2)\wedge d\phi_i
\wedge dx^+\wedge d^2z\,.
\eea
The $\mu_i$ are parameterised as
\be
\mu_1=\sin\theta\,, \qquad \mu_2=\cos\theta\,\sin\psi\,, \qquad
\mu_3=\cos\theta\,\cos\psi\,,
\ee
in terms of the angles on a $2$-sphere.

\subsection{$D=6$ oxidised to Romans massive theory}
The bosonic sector of Romans massive theory \cite{rom2} is described 
by the Lagrangian
\bea
&&{\cal L}_{10}=\hat R\hat\ast\oneone-\ft12\hat\ast d\hat\phi
\wedge d\hat\phi - \ft12e^{\ft32\hat\phi}\hat\ast \hat F_\2
\wedge\hat F_\2 - \ft12e^{-\hat\phi}
\hat\ast\hat F_\3\wedge \hat F_\3
-\ft12e^{\ft12\hat\phi}\hat\ast \hat F_\4\wedge \hat F_\4\nn\\
&&\quad - \ft12d\hat A_\3\wedge d\hat A_\3\wedge\hat A_\2 
- \ft16m\, d\hat A_\3\wedge(\hat A_\2)^3 
- \ft1{40}m^2(\hat A_\2)^5 
-\ft12m^2e^{\ft52\hat\phi}\hat\ast\oneone\,,
\eea
where the field strengths are defined as
\bea
&&\hat F_\2=d\hat A_\1 + m\hat A_\2\,, \qquad 
\hat F_\3=d\hat A_\2\,,\nn\\
&&\hat F_\4=d\hat A_\3 + \hat A_\1\wedge d\hat A_\2 
+ \ft12m\hat A_\2\wedge\hat A_\2\,. 
\eea
Note that the Bianchi identities in this theory are given by
\be
d\hat F_\4=\hat F_\2\wedge\hat F_\3\,, \qquad d\hat F_\3=0\,, 
\qquad d\hat F_\2=m\hat F_\3\,.
\ee
Using the embedding formulae obtained in \cite{clp5} we can lift our
six dimensional solution to a solution of the above theory. It is
given by (with $m=g$)
\bea
d\hat s_{10}^2\!\!\!&=&\!\!\!s^{\ft1{12}}\,[ds_6^2+\ft49g^{-2}d\xi^2
+ \ft19g^{-2}c^2\,(\sigma_1^2 + \sigma_2^2 + h_3^2)]\,,\nn\\
\hat F_\4\!\!\!&=&\!\!\!
\ft{10}{81}g^{-3}s^{1/3}c^3\,d\xi\wedge\epsilon_\3
- \ft2{9\sqrt2} g^{-2}e^{-3g\rho}S[s^{1/3}c\,\sigma_3\wedge d\xi
- \ft12s^{4/3}c^2\,\sigma_1\wedge\sigma_2]\wedge dx^+\wedge d\rho\,,\nn\\
\hat F_\3\!\!\!&=&\!\!\! 0\,,\qquad
\hat F_\2= 0\,,\qquad e^{\hat\phi}=s^{-5/6},
\eea
where $ds_6^2$ is given by (\ref{ansatz6}) and (\ref{h6}), and
$s, c, \epsilon_\3$ and $\sigma_i$ have the same definitions
as before and $h_3=\sigma_3-\ft1{\sqrt2}S(1-e^{-3g\rho})dx^+$.

\subsection{$D=7$ oxidised to $D=11$}
Using the embedding ansatz in \cite{lp} we obtain
\bea
d\hat s_{11}&=&ds_7^2+\ft14g^{-2}d\xi^2+\ft1{16}g^{-2}c^2
(\sigma_1^2 + \sigma_2^2 + h_3^2)\,,\\\
\hat A_\3&=&\ft1{64}g^{-3}(2s+s\,c^2)\epsilon_\3
+ \ft1{8\sqrt2} g^{-2}S\,e^{-4g\rho}s\, 
dx^+\wedge d\rho\wedge\sigma_3\,,\nn
\eea
where $ds_7^2$ is given by (\ref{ansatz7}) and (\ref{h7}). 
The field strength $\hat F_\4=d\hat A_\3$ is
\bea
\hat F_\4&=&\ft3{64}g^{-3}c^3\,d\xi\wedge \epsilon_\3
+ \ft1{8\sqrt2} g^{-2}S\,e^{-4g\rho}c\,
dx^+\wedge d\rho\wedge d\xi\wedge \sigma_3\nn\\
&&\qquad + \ft1{16\sqrt2} g^{-2}S\,e^{-4g\rho}\,s\,c^2\,
dx^+\wedge d\rho\wedge \sigma_1\wedge \sigma_2\,,
\eea
where $s,c,\epsilon_\3$ and $\sigma_i$ have the same definitions
as before and $h_3=\sigma_3-\ft1{\sqrt2}S(1-e^{-4g\rho})dx^+$.

\section{A general class of pp-waves}

In this appendix we present the AdS pp-waves supported by an arbitrary
$n$-form field strength in any dimensions $D$. The Lagrangian for such
a system is given by
\be
e^{-1}{\cal L}=R-\fft1{2n!}F_{(n)}^2+(D-1)(D-2)g^2,
\ee
where the field strength is defined as $F_{(n)}=dA_{(n-1)}$.
Our pp-wave ansatz is
\bea
ds^2 &=& e^{2g\rho} (-4dx^+\, dx^- + H\, (dx^+)^2 + dz^2) +
d\rho^2,\nn\\
A_{(n-1)}&=&{\Big(} zS_1(x^+)-\fft{S_2(x^+)}{g(D-2n+1)}
(e^{-(D-2n+1)g\rho}-1){\Big)}\, dx^+ \wedge d^{\,n-2}z\,.
\eea
The field strength and its dual are
\bea
F_{(n)}&=&-S_1\, dx^+ \wedge dz^{n-1} + S_2 e^{-(D-2n+1)g\rho}
d\rho \wedge dx^+ \wedge d^{n-2}z\,,\nn\\
\ast F_{(n)}&=&S_1 e^{(D-2n-1)g\rho}d\rho
\wedge dx^+ \wedge d^{\,D-n-2}z - S_2 dx^+ \wedge d^{\,D-n-1}z\,.
\eea
Thus the equation of motion $d{*F_{(n)}}=0$ is trivially satisfied.
The Einstein equation implies
\bea
\square H &=& -S_1^2e^{-2ng\rho} - S_2^2e^{-2(D-n)g\rho}\,,\nn\\
\square&=&\del_\rho^2+g(D-1)\del_\rho
+e^{-2g\rho}\sum_{i=1}^{D-3}\del_i^2\,,
\eea
with the solution given by
\bea
H(x^+,\rho,z_i)&=&a + b\,e^{-(D-1)g\rho} + \fft{e^{-2g\rho}}{2g^2(D-3)}
\sum_{i=1}^{D-3}c_i + \fft{S_1^2\, e^{-2ng\rho}}{2g^2n(D-2n-1)}\nn\\
&-&\fft{S_2^2\, e^{-2(D-n)g\rho}}{2g^2(D-n)(D-2n+1)}
+ \ft12\sum_{i=1}^{D-3}c_iz_i^2\,.
\label{ndppwave1}
\eea
The $a$, $b$ and $c_i$ are functions of $x^+$.
This solution is not valid for $D=2n-1$ or $D=2n+1$ which have to 
be considered separately. We find that
\bea
H(D=2n+1,x^+,\rho,z_i)&=&a + b\, e^{-2ng\rho} + \fft{e^{-2g\rho}}{4g^2(n-1)}
\sum_{i=1}^{2(n-1)}\!\! c_i + \fft{(2ng\rho+1)S_1^2}{4n^2g^2}\, 
e^{-2ng\rho}\nn\\
&-&\fft{S_2^2}{4g^2(n+1)}\, e^{-2(n+1)g\rho}
+ \ft12 \!\! \sum_{i=1}^{2(n-1)} \!\! c_i z_i^2\,,
\label{ndppwave2}
\eea
and $H(D=2n-1)$ can be obtained from $H(D=2n+1)$ by making
the substitution
\be
n \rightarrow n-1 \qquad \mbox{and} 
\qquad S_1 \longleftrightarrow S_2\,.
\ee
(This substitution is not performed on the field strength.)


\begin{thebibliography}{99}

\bm{bfhp}
M. Blau, J. Figueroa-O'Farrill, C. Hull and G. Papadopoulos,
{\it ``A new maximally supersymmetric background of IIB superstring theory,}
JHEP {\bf 0201}, 047 (2002), hep-th/0110242.

\bm{kg} J. Kowalski-Glikman, {\it Vacuum states in supersymmetric
 Kaluza-Klein theory}, Phys. Lett. {\bf B134}, 194 (1984).

\bm{bmn}
D. Berenstein, J.M. Maldacena and H. Nastase,
{\it Strings in flat space and pp-waves from $N = 4$ super Yang Mills,}
JHEP {\bf 0204}, 013 (2002), hep-th/0202021.

\bm{malda} J.M. Maldacena, {\it The large N limit of
superconformal field theories and supergravity}, Adv. Theor. Math.
Phys. 2 231 (1998); Int. J. Theor. Phys. {\bf 38} 1113 (1999),
hep-th/9711200.

\bm{gkp} S.S. Gubser, I.R. Klebanov and A.M. Polyakov, {\it Gauge
theory correlators from non-critical string theory}, Phys.\ Lett.\
{\bf B428} 105 (1998), hep-th/9802109.

\bm{wit} E.~Witten, {\it Anti-de Sitter space and holography},
Adv.\ Theor.\ Math.\ Phys.\  {\bf 2} 253 (1998), hep-th/9802150.

\bm{metsaev}R.R. Metsaev,
{\it Type IIB Green-Schwarz superstring in plane wave Ramond-Ramond 
background,}
Nucl.\ Phys.\ {\bf B625}, 70 (2002), hep-th/0112044.

\bm{clp1} M. Cveti\v c, H. L\"u and C.N. Pope,
{\it Penrose limits, pp-waves and deformed M2-branes,}
Phys.\ Rev.\ {\bf D69}, 046003 (2004), hep-th/0203082.

\bm{clp2} M. Cveti\v c, H. L\"u and C.N. Pope,
{\it M-theory pp-waves, Penrose limits and supernumerary supersymmetries,}
Nucl.\ Phys.\ B {\bf 644}, 65 (2002), hep-th/0203229.

\bm{gh} J.P. Gauntlett and C.M. Hull,
{\it pp-waves in 11-dimensions with extra supersymmetry,}
JHEP {\bf 0206}, 013 (2002), hep-th/0203255.

\bm{bena} I. Bena and R. Roiban, {\it Supergravity pp-wave solutions 
with 28 and 24 supercharges,} Phys.\ Rev.\ {\bf D67}, 125014 (2003),
hep-th/0206195.

\bibitem{mich1} J. Michelson,
{\it (Twisted) toroidal compactification of pp-waves,}
Phys.\ Rev.\ {\bf D66}, 066002 (2002), hep-th/0203140.

\bm{mich2} J. Michelson, {\it A pp-wave with 26 supercharges,}
Class.\ Quant.\ Grav.\  {\bf 19}, 5935 (2002),
hep-th/0206204.

\bm{lpv} H. L\"u and J.F. V\'azquez-Poritz,
{\it Penrose limits of non-standard brane intersections,}
Class.\ Quant.\ Grav.\  {\bf 19}, 4059 (2002), hep-th/0204001.

\bm{clpboost} M. Cveti\v c, H. L\"u and C.N. Pope,
{\it Spacetimes of boosted p-branes, and CFT in infinite-momentum frame,}
Nucl.\ Phys.\ {\bf B545}, 309 (1999), hep-th/9810123.

\bibitem{bcr} D. Brecher, A. Chamblin and H.S. Reall,
{\it AdS/CFT in the infinite momentum frame,}
Nucl.\ Phys.\ {\bf B607}, 155 (2001), hep-th/0012076.

\bm{kaig} V.R. Kaigorodov, {\it Einstein spaces of maximum mobility},
Dokl. Akad. Nauk. SSSR {\bf 146} 793 (1962); Sov. Phys.  Doklady 
{\bf 7}, 893 (1963).

\bm{siklos} S.T.C. Siklos, {\it Lobatchevski plane gravitational waves,}
in Galaxies, Axisymmetric Systems and Relativity, ed. M.A.H. MacCallum,
Cambridge University Press, Cambridge, England, 1985.
                                                                                
\bm{ozsvath} I. Ozsvath, I. Robinson and K. Rozga,
{\it Plane-fronted gravitational and electromagnetic waves in spaces
with cosmological constant,} J. Math.\ Phys.\ {\bf 26} (1985) 1755.
                                                                                
\bm{gibbonsruback} G.W. Gibbons and P.J. Ruback, {\it Classical gravitons and
their stability in higher dimensions,} Phys.\ Lett.\ {\bf B171} (1986) 390.

\bm{freed} D.Z. Freedman and A. Das, {\it Gauge internal symmetry in
extended supergravity,} Nucl.\ Phys.\ {\bf B120}, 221 (1977).

\bm{frad} E.S. Fradkin and M.A Vasiliev, {\it Model of supergravity
with minimal electromagnetic interaction,} Lebedev Institute preprint
N 197 (1976).

\bm{cai} R.G. Cai, {\it Boosted domain wall and charged Kaigorodov
space,} Phys.\ Lett.\ {\bf B572}, 75 (2003), hep-th/0306140.

\bm{klemm1} M.M. Caldarelli and D. Klemm, {\it All supersymmetric
solutions of $N = 2$, $D = 4$ gauged supergravity,} JHEP {\bf 0309},
019 (2003), hep-th/0307022.

\bm{klemm2} S.L. Cacciatori, M.M. Caldarelli, D. Klemm and D.S.
Mansi, {\it More on BPS solutions of $N = 2$, $D = 4$ gauged
supergravity,} JHEP {\bf 0407}, 061 (2004), hep-th/0406238.

\bm{mg} M. G\"unaydin, G. Sierra and P.K. Townsend, {\it The geometry
of N=2 Mawxwell-Einstein supergravity and Jordan algebras,} Nucl.\
Phys.\ {\bf B242}, 244 (1984).

\bm{gg} J.P. Gauntlett and J.B. Gutowski,
{\it All supersymmetric solutions of minimal gauged supergravity in five
dimensions,} Phys.\ Rev.\ {\bf D68}, 105009 (2003), 
erratum-ibid. {\bf D70}, 089901 (2004), hep-th/0304064.

\bm{rom1} L.J. Romans, {\it The F(4) gauged supergravity in
six-dimensions,} Nucl.\ Phys.\ {\bf B269}, 691 (1986).

\bm{clp5} M. Cveti\v c, H. L\"u and C.N. Pope, {\it Gauged
six-dimensional supergravity from massive type IIA,} Phys.\ Rev.\
Lett.\ {\bf 83}, 5226 (1999), hep-th/9906221.

\bm{radion} J.T. Liu, H. L\"u and C.N. Pope,
{\it The radion mode in consistent brane-world reductions,}
hep-th/0212037.

\bm{ker} J. Kerimo, J.T. Liu, H. L\"u and C.N. Pope, {\it Variant
${\cal N}=(1,1)$ supergravity and (Minkowski)$_4\times S^2$ vacua,}
Class.\ Quant.\ Grav. {\bf 21}, 3287 (2004), hep-th/0401001.

\bm{town} P.K. Townsend and P. van Nieuwenhuizen, {\it Gauged seven
dimensional supergravity,} Phys.\ Lett.\ {\bf 125B}, 41 (1983).

\bm{lp} H. L\"u and C.N Pope, {\it Exact embedding of $N=1,D=7$ gauged
supergravity in $D=11$,} Phys.\ Lett.\ {\bf B467}, 67 (1999),
hep-th/9906168.

\bm{clp3} M. Cveti\v c, M.J. Duff, P. Hoxa, J.T. Liu, H. L\"u, J.X. Lu,
R. Martinez-Acosta, C.N. Pope, H. Sati and T.A. Tran,
{\it Embedding AdS black holes in ten and eleven dimensions\,}
Nucl.\ Phys.\ {\bf B558}, 96 (1999), hep-th/9903214.

\bm{clp4} M. Cveti\v c, H. L\"u and C.N. Pope, {\it Four-dimensional
N$=4$, $SO(4)$ gauged supergravity from $D=11$\,} 
Nucl.\ Phys.\ {\bf B574}, 761 (2000), hep-th/9910252.

\bm{kumar} A. Kumar and H.K. Kunduri,
{\it Gravitational wave solutions in string and M-theory AdS backgrounds,}
hep-th/0405261.

\bm{tuan} H. L\"u, C.N. Pope and T.A. Tran,
{\it Five-dimensional ${\cal N} = 4$, $SU(2) \times U(1)$
gauged supergravity from type IIB,} Phys.\ Lett.\ {\bf B475},
261 (2000), hep-th/9909203.

\bm{rom2} L.J. Romans, {\it Massive N=2a supergravity in ten dimensions,}
Phys.\ Lett.\ {\bf B169}, 374 (1986).

\end{thebibliography}
\end{document}